\documentclass[useAMS,usenatbib]{mn2e}

\usepackage{graphicx}
\usepackage{times}%
\usepackage{natbib}
\usepackage{graphics}
\usepackage{epsfig}
\usepackage{array}
\usepackage{stfloats}
\usepackage{fixltx2e}
\usepackage{amsmath}

\newcommand{\gtrsim}{\ga}
\newcommand{\lesssim}{\la} 

\newcommand{\gsim}{\,\lower2truept\hbox{${>\atop\hbox{\raise4truept\hbox{$\sim$}}}$}\,}

\newcommand{\be}{\begin{equation}}
\newcommand{\ee}{\end{equation}}
\newcommand{\bea}{\begin{eqnarray}}
\newcommand{\eea}{\end{eqnarray}}

\def\ltsima{$\; \buildrel < \over \sim \;$}
\def\simlt{\lower.5ex\hbox{\ltsima}}
\def\gtsima{$\; \buildrel > \over \sim \;$}
\def\simgt{\lower.5ex\hbox{\gtsima}}

\title[]{Structure formation in Multiple Dark Matter cosmologies with long-range scalar interactions}

\author[M. Baldi]{Marco Baldi\thanks{Email: mail@marcobaldi.it}
\\Excellence Cluster Universe, Boltzmannstr.~2, D-85748 Garching, Germany
\\University Observatory, Ludwig-Maximillians University Munich, Scheinerstr. 1, D-81679 Munich, Germany}

\begin{document}

%\date{Accepted ???. Received ???; in original form }

\pagerange{\pageref{firstpage}--\pageref{lastpage}} \pubyear{2011}

\maketitle

\label{firstpage}

\begin{abstract}	
\ \\
Long-range scalar forces with gravitational strength can be ruled out by accurate tests
of scalar-tensor theories in the solar system. However, such tests are based on the motion of celestial bodies
made of standard baryonic particles, which leaves room for possible scalar interactions
in the dark sector of the universe if a new scalar degree of freedom couples only to dark matter particles.
In particular, an interaction between Cold Dark Matter (CDM) and a classical scalar field playing the
role of the cosmic dark energy (DE) might provide such long-range dark interactions without conflicting with solar system bounds.
Although presently available observations allow to constrain such interactions
to a few percent of the gravitational strength, some recent studies have
shown that if CDM is composed by two different particle species having opposite couplings
to the DE field, such tight constraints can be considerably relaxed, allowing for long-range
scalar forces of order gravity without significantly affecting observations both at the background 
and at the linear perturbations level. In the present work, we extend the investigation of such Multiple Dark Matter scenarios
to the nonlinear regime of structure formation, by presenting the first N-body simulations ever performed
for these cosmologies. Our results highlight some characteristic footprints of long-range scalar forces that 
arise only in the nonlinear regime for specific models that would be otherwise practically indistinguishable from
the standard $\Lambda $CDM scenario both in the background and in the growth of linear density perturbations. Among these effects, the formation of 
{\em ``mirror"} cosmic structures in the two CDM species, the suppression of the nonlinear matter power spectrum at $k \gtrsim 1\, h/$Mpc, and the fragmentation of collapsed halos, represent
peculiar features that might provide a direct way to constrain this class cosmological models.

\end{abstract}

\begin{keywords}
dark energy -- dark matter --  cosmology: theory -- galaxies: formation
\end{keywords}

%*****************************************************************************

\section{Introduction}
\label{i}

While the existence of two distinct forms of gravitating energy in the Universe which do not
interact with the elecromagnetic field -- named Dark Matter and Dark Energy -- is now supported by a large number of 
independent and complementary observations \citep[see e.g.][and references therein]{Bertone_Hooper_Silk_2005,Bartelmann_2010,Astier_Pain_2012}, their fundamental nature is still completely
unknown and represents the main focus of most of the present research efforts in the fields of theoretical and observational
cosmology and astroparticle physics. 
Most of our knowledge about the phenomenology of Dark Energy (DE) and Cold Dark Matter (CDM) comes from the
direct observation of their gravitational effects on the evolution of cosmic structures and on the global evolution of the Universe
as a whole \citep[see e.g.][]{Riess_etal_1998,Schmidt_etal_1998,Perlmutter_etal_1999,Percival_etal_2001,SNLS,Giannantonio_etal_2008, Guzzo_etal_2008,Fu_etal_2008,Mantz_etal_2010, Schrabback_etal_2010,Percival_etal_2010,Sherwin_etal_2011}. However, such effects do not allow to drive clear conclusions about the fundamental constituents of these two
dark components, and in particular do not allow to resolve a possible internal complexity of each of these fluids. 
Furthermore, 
all the experimental and observational initiatives aimed at a direct detection of CDM particles in the laboratory have so far
failed to provide an unambiguous and statistically significant identification of any of the proposed CDM particle candidates \citep[for an excellent recent review see e.g.][]{Bergstrom_2012}.

Such a deep uncertainty on the fundamental nature of these two cosmic fluids leaves room for theoretical speculations about a possible hidden complexity of the dark sector, as e.g. a dynamical nature of DE based on the evolution of an ubiquitous scalar field as in the case of {\em Quintessence} \citep[][]{Wetterich_1988,Ratra_Peebles_1988} or {\em k-essence} \citep[][]{kessence}, a
direct interaction between DE and CDM \citep[see e.g.][]{Wetterich_1995,Amendola_2000,Farrar2004,Pettorino_Baccigalupi_2008,Baldi_2011a}, a multi-particle composition of the CDM fluid \citep[see e.g.][]{Khlopov_1995,Brookfield_VanDeBruck_Hall_2008,Baldi_2012a,Maccio_etal_2012}, or the presence of long-range scalar or vector interactions between CDM particles \citep[see e.g.][]{Hellwing_etal_2010, Keselman_Nusser_Peebles_2010,Loeb_Weiner_2011,Aarssen_Bringmann_Pfrommer_2012}.
The investigation of such theoretical scenarios in the last years allowed to put tight constraints on their characteristic parameters
and to strongly limit the range of viable theoretical extensions of the standard $\Lambda $CDM cosmological model. In particular, 
models featuring a direct interaction between a DE scalar field and a single family of CDM particles have been widely studied for what concerns their impact on the background expansion history of the universe, as
well as for their effects on the growth of density perturbations in the linear \citep[e.g.][]{Bean_etal_2008,LaVacca_etal_2009,Xia_2009} and nonlinear \citep{Maccio_etal_2004,Baldi_etal_2010,Baldi_2011b,Li_Barrow_2011} regimes of structure formation, which allowed
to put relatively tight constraints on the coupling constant between DE and CDM and consequently on the strength of its associated long-range
scalar force. 

Nevertheless, the recent works of \citet{Brookfield_VanDeBruck_Hall_2008} and \citet{Baldi_2012a} have shown
how in the presence of a higher level of internal complexity of the CDM sector, as e.g. in the case of multiple CDM particle species
with different couplings to a DE scalar field, the background and linear constraints on such dark scalar forces might be significantly 
relaxed, even for scenarios that do not introduce any additional parameters with respect to standard models of interacting DE. 
In the present work, we will extend the investigation of these coupled DE models with Multiple Dark Matter families to the
nonlinear regime of structure formation by means of a series of intermediate-resolution N-body simulations, with the aim to
identify possible characteristic footprints capable to constrain the parameter space of such models beyond the very loose
bounds provided by the background expansion history and by the linear growth of structures. As our study will show, although nonlinear structures result substantially affected in these models for sufficiently large values of the DE-CDM coupling constant,
a long-range scalar interaction with a strength comparable to that of standard gravity cannot be easily ruled out even in the nonlinear
regime. Such results open new possible directions for phenomenological extensions of the standard cosmological model, allowing
for new long-range interactions of gravitational strength. The present work therefore represents a rather qualitative investigation of the effects of multiple interactions between DE and CDM in the nonlinear regime, and is mainly aimed at a broad sampling of the parameter space of these scenarios which could provide a useful guideline to drive future numerical investigations towards selected parameter combinations.

Our paper is organized as follows. In Section~\ref{MDM} we will introduce the specific cosmological models with Multiple Dark Matter
families interacting with DE. In Section~\ref{sims} we will describe our set of numerical simulations and in Section~\ref{Results} we will present the main results
of our analysis, concerning the shape of large-scale structures and the nonlinear matter power spectrum in such cosmological models. 
Finally, in Section~\ref{conclusions} we will draw our conclusions.

\section{Multiple Dark Matter models}
\label{MDM}

If the observed accelerated expansion of the Universe is sourced by some dynamical scalar
field such as {\em Quintessence} \citep[][]{Wetterich_1988,Ratra_Peebles_1988} or {\em k-essence}
\citep[][]{kessence} rather than by a simple cosmological constant, it is natural to consider the possible
direct interactions that such DE component might have with the other fields of the dark and of the visible sector.
Any such interaction would then mediate a new scalar force between massive particles, and in most viable scenarios this force 
is long-ranged since the DE field has to be sufficiently light in order to explain the energy scale of DE, and its
propagation length is therefore correspondingly large, generically comparable to the cosmic horizon \citep[for a general review of interacting DE models, see Chapter 1 of][]{Euclid_TWG}.

Universal interactions between a cosmic scalar and all massive particle fields in the universe constitute the
broad family of scalar-tensor theories of gravity and Extended Quintessence models \citep[][]{Brans_Dicke_1961,Appleby_Weller_2010,Baccigalupi_Matarrese_2000,Perrotta_etal_2000,Pettorino_etal_2005}, and their interaction strength is very tightly constrained by solar system
tests of General Relativity \citep[see e.g.][]{Bertotti_Iess_Tortora_2003,Will_2005} unless some screening mechanism -- as e.g. the Chameleon \citep[][]{Khoury_Weltman_2004}, the Vainshtein \citep[][]{Vainshtein_1972,Deffayet_etal_2002}, or the Symmetron \citep[][]{Hinterbichler_Khoury_2010} effects -- hides the scalar force in our local environment. However, such tight constraints
do not apply to a selective interaction between a DE scalar field and the dark matter sector, as first proposed by \citet{Damour_Gibbons_Gundlach_1990}. Such possibility is the basic ingredient of coupled DE scenarios \citep[see e.g.][]{Wetterich_1995,Amendola_2000,Amendola_2004,Pettorino_Baccigalupi_2008,Baldi_2011a,Baldi_2011c} that have been very widely investigated in the last decade as they could provide dynamical solutions to the fine-tuning problems of the cosmological constant. In these models, DE and CDM (which is assumed to be made of a single species of weakly interacting and massive fundamental particles) interact through the exchange of energy-momentum such that the total stress-energy tensor of the Universe is still conserved while the individual fluids feature an additional source term in their continuity equations:
\begin{eqnarray}
& &\nabla _{\mu}T^{\mu }_{\nu } = 
%\nabla _{\mu}\left[ T^{\mu ({\rm DE})}_{\nu } + T^{\mu ({\rm CDM})}_{\nu }\right] =
0\,; \\
& &\nabla _{\mu}T^{\mu ({\rm DE})}_{\nu } = - \nabla _{\mu}T^{\mu ({\rm CDM})}_{\nu } = CT^{\mu ({\rm CDM})}_{\mu }\nabla _{\nu }\phi \,,
\end{eqnarray}
where $C$ is a constant and $\phi $ is the homogeneous value of the DE scalar field.
These scenarios have been studied thoroughly concerning their effects on the background expansion history of the universe \citep[][]{Amendola_2000,Amendola_2004,CalderaCabral_2009,Clemson_etal_2011}, their
impact on the growth of linear density perturbations \citep[][]{Amendola_Quercellini_2003,Pettorino_Baccigalupi_2008,Koyama_etal_2009,LaVacca_etal_2009,Amendola_Pettorino_Quercellini_2012}, and their effects on the nonlinear dynamics of collapsed structures \citep[][]{Maccio_etal_2004,Baldi_etal_2010,Li_Barrow_2011,Baldi_2011b,CoDECS}. Such investigations have provided relatively tight constraints on the maximum allowed values of the interaction constant $C$ based on a wide range of observational probes \citep[see e.g.][]{Bean_etal_2008,Baldi_Pettorino_2011,Baldi_Viel_2010,Xia_2009}. 

It goes beyond the scope of the present paper to provide a comprehensive review of the basic properties and the main observational constraints of coupled DE models, and we refer the interested reader to the mentioned literature for a more detailed discussion of these issues. Here instead we want to focus on the fact that present constraints seem to exclude with high statistical confidence the possibility of couplings with strength comparable to the gravitational interaction. If we define, following the literature \citep[e.g.][]{Amendola_2000} the dimensionless coupling $\beta $ through the relation:
\begin{equation}
C\equiv \sqrt{\frac{2}{3}}\frac{1}{M_{\rm Pl}}\beta \,,
\end{equation}
where $M_{\rm Pl} \equiv 1/\sqrt{8\pi G}$ is the reduced Planck mass with $G$ the Newton's constant, present observational bounds restrict the coupling to $\beta \lesssim 0.15$ at most \citep[see e.g.][]{Bean_etal_2008,Xia_2009,Baldi_Viel_2010,LaVacca_etal_2009}, corresponding to a scalar interaction constant $4G\beta ^{2}/3\approx 0.03G$
 \citep[see][]{Amendola_2004} with a few percent of the gravitational strength. 

All these observational constraints apply to standard coupled DE models, where the coupling $C$ is a constant and where only one
CDM particle species is present in the universe. More complex scenarios, featuring e.g. a time evolution of the coupling \citep[][]{Baldi_2011a}, a ``bouncing" DE self-interaction potential \citep[as e.g. in][]{Baldi_2011c} or the existence of more than one CDM species interacting with DE \citep[][]{Brookfield_VanDeBruck_Hall_2008,Baldi_2012a} have been proposed in recent years and have been shown to easily evade such constraints, allowing for larger values of the coupling. In particular, \citet{Brookfield_VanDeBruck_Hall_2008} investigated a general framework where multiple CDM fluids interact with individual couplings to the DE scalar field, highlighting how the presence of multiple dark matter species determines new critical points of the background dynamical system for which the total effective background coupling is suppressed, thereby reducing the impact of the coupling on the cosmic expansion history. More recently, \citet{Baldi_2012a} studied in large detail a specific realization of such general framework, where CDM is composed by only two different particle species completely degenerate in all their physical properties except for the sign of their interaction constant with the DE scalar field. Such specific Multiple Dark Matter coupled DE model (or Multi-coupled DE, McDE hereafter) is described at the background level by the following system of dynamic equations:
\begin{eqnarray}
\label{klein_gordon}
\ddot{\phi } + 3H\dot{\phi } + \frac{dV}{d\phi } &=& +C \rho _{+} - C \rho _{-}\,, \\
\label{continuity_plus}
\dot{\rho }_{+} + 3H\rho _{+} &=& -C \dot{\phi }\rho _{+} \,, \\
\label{continuity_minus}
\dot{\rho }_{-} + 3H\rho _{-} &=& +C \dot{\phi }\rho _{-} \,, \\
\label{continuity_radiation}
\dot{\rho }_{r} + 4H\rho _{r} &=& 0\,, \\
\label{friedmann}
3H^{2} &=& \frac{1}{M_{{\rm Pl}}^{2}}\left( \rho _{r} + \rho _{+} + \rho _{-} + \rho _{\phi} \right)\,,
\end{eqnarray}
where an overdot represents a derivative with respect to the cosmic time $t$, $H\equiv \dot{a}/a$ is the Hubble function,
the two CDM species are denoted by the subscripts ``$+$" and ``$-$" according to the sign of their coupling constant, and the total CDM density is given by $\rho _{\rm CDM} = \rho _{+}+\rho _{-}$. 

In the present paper, we will focus on this particular class of models which has the appealing feature of requiring no additional free parameters with respect to a standard coupled DE scenario with constant coupling.
Without loss of generality for the purposes of our discussion, we will not include in the analysis the uncoupled baryonic component of the universe, and we will always assume an exponential form for the scalar self-interaction potential $V(\phi )$ appearing in the Klein-Gordon equation (\ref{klein_gordon}).
The background dynamics of this scenario has been described in detail in \citet{Baldi_2012a} which showed how for reasonable initial conditions (i.e. without a too large asymmetry between the two different CDM species in the early universe) the expansion
history of the model is completely indistinguishable from $\Lambda $CDM even for dimensionless couplings as large as $\beta = 10$. This is due to the fact that during matter domination the scalar field is trapped in the minimum of its effective potential $V_{\rm eff}$  defined by:
\begin{equation}
\frac{d V_{\rm eff}}{d\phi }\equiv \frac{dV}{d\phi} -C\rho _{+} + C\rho _{-}\,,
\end{equation}
and the scalar field velocity $\dot{\phi }$ is consequently very small, thereby suppressing the effect of the interaction in Eqs.~(\ref{continuity_plus},\ref{continuity_minus}). 

\begin{figure}
\includegraphics[scale=0.45]{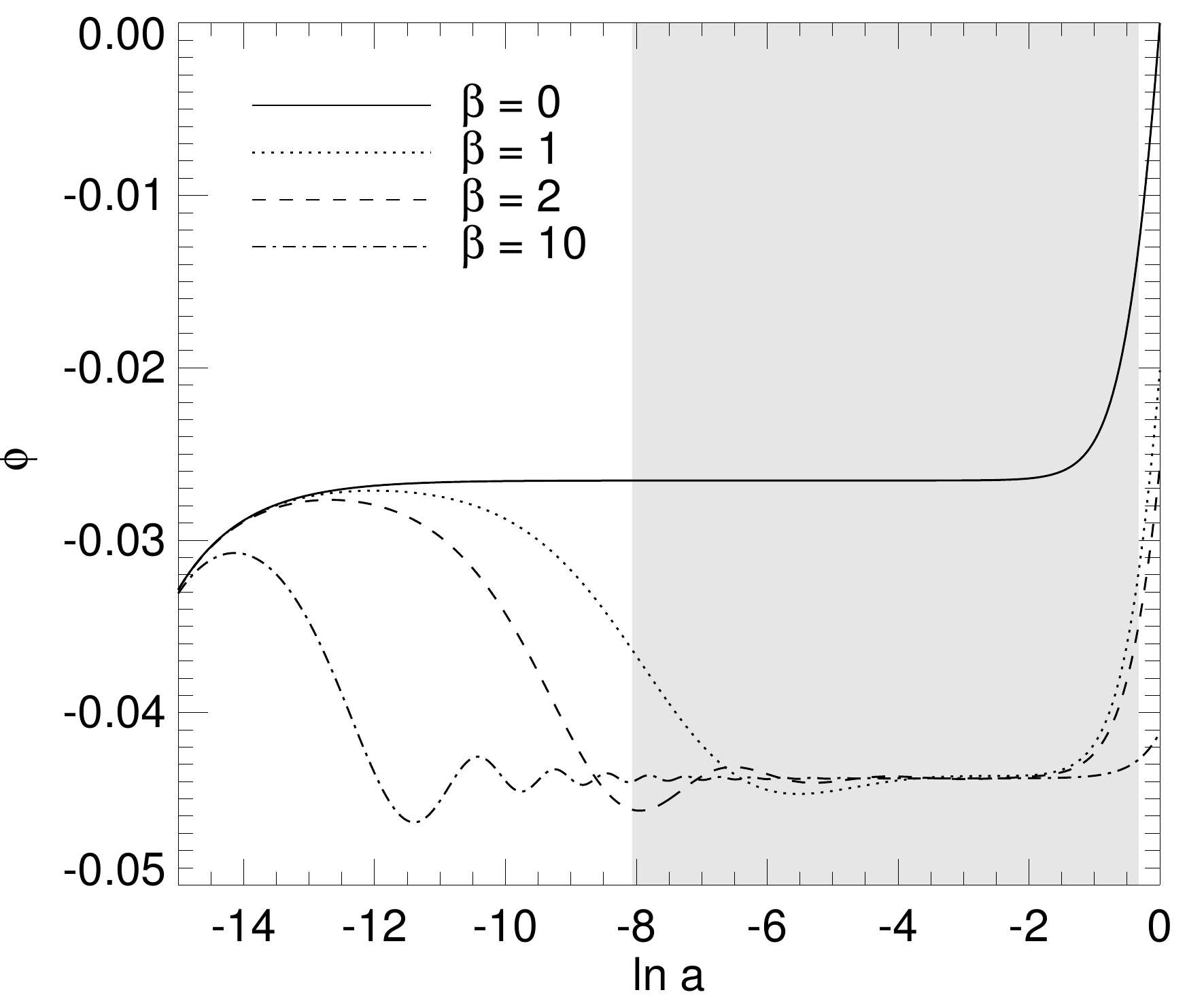}
\caption{The evolution of the DE scalar field $\phi $ as a function of the e-folding time $\ln a$ in an uncoupled scenario (solid) and in McDE cosmologies with different values of the coupling $\beta $ (dotted, dashed, and dot-dashed). As one can see from the figure, after the initial evolution the scalar field gets trapped in the minimum of its effective self-interaction potential during most of matter domination (represented by the grey-shaded area) and during this period behaves like an uncoupled field.}
\label{fig:phi}
\end{figure}
\normalsize

In Fig.~\ref{fig:phi} we show the evolution of the scalar field $\phi $ as a function of the e-folding time $\ln a$ for a range of McDE models with different couplings starting from the same initial conditions at high redshift. As the plot shows, even for coupling values larger than unity (dotted, dashed, and dot-dashed curves) the scalar field is attracted to a fixed point where it remains constant during the whole matter dominated epoch, identified in the plot by the grey-shaded area, thereby behaving like an uncoupled system (solid curve). In other words, the background evolution of the universe can be completely screened from arbitrarily large coupling values if two CDM species with an opposite sign of the coupling both interact with the DE scalar field. Such screening mechanism therefore does not allow to distinguish a McDE model from an uncoupled cosmology -- or even from $\Lambda $CDM -- through geometrical probes of the cosmic evolution.
\ \\

However, \cite{Brookfield_VanDeBruck_Hall_2008} already showed that this degeneracy can be easily broken by the evolution of linear
density perturbations. In fact, even for a vanishing effective coupling in the background equations~(\ref{klein_gordon}-\ref{friedmann}), linear density perturbations feature a significantly enhanced growth with respect to $\Lambda $CDM in the presence of such large coupling values. This is due to the fact that the symmetry between the two CDM species that holds in the background as a consequence of the new matter-dominated critical point, is broken at the level of density perturbations by any oscillation of the field around the minimum of the effective potential, which determines a different friction term for the fluctuations in the two different CDM species \citep[see][]{Baldi_2012a}. This can be understood by having a look at the linear perturbations equations for the specific case of two CDM species with opposite couplings:
\begin{eqnarray}
\label{gf_plus}
\ddot{\delta }_{+} = -2H\left[ 1 - \beta \frac{\dot{\phi }}{H\sqrt{6}}\right] \dot{\delta }_{+} + 4\pi G \left[ \rho _{-}\delta _{-} \Gamma_{R} + \rho _{+}\delta _{+}\Gamma_{A}\right] \,, \\
\label{gf_minus}
\ddot{\delta }_{-} = -2H\left[ 1 + \beta \frac{\dot{\phi }}{H\sqrt{6}}\right] \dot{\delta }_{-} + 4\pi G \left[ \rho _{-}\delta _{-} \Gamma _{A} + \rho _{+}\delta _{+}\Gamma_{R}\right]\,.
\end{eqnarray}
In Eqs.~(\ref{gf_plus},\ref{gf_minus}) the $\Gamma $ factors are defined as:
\begin{equation}
\label{def_gamma}
\Gamma _{A} \equiv 1 + \frac{4}{3}\beta ^{2}\,, \quad \Gamma _{R}\equiv 1 - \frac{4}{3}\beta ^{2} \,,
\end{equation}
and represent attractive ($\Gamma _{A}$) or repulsive ($\Gamma _{R}$) corrections to gravity due to the
long-range fifth-force mediated by the DE scalar field, while the second terms in the first squared brackets on the right-hand side are the friction terms associated to momentum conservation, that break the symmetry of the two equations. From Eqs.~(\ref{gf_plus}-\ref{def_gamma}) it is then clear that any dynamics of the scalar field (i.e. $\dot{\phi }\neq 0$) will induce a different growth of the density fluctuations in the two CDM species, which for couplings larger than $\beta = \sqrt{3}/2$ will then start repelling each other due to their mutual repulsive scalar force \citep[see][for a more detailed discussion on the origin of the repulsion between fluctuations in the two different species]{Baldi_2012a}. As a result, the perturbation in the more overdense species will keep growing while the other will start decaying, thereby producing isocurvature modes even when starting from purely adiabatic initial conditions. In any case, the growth of overdensities in one of the CDM species will always be faster than the depletion of the corresponding underdensities in the other species generated by the repulsive scalar force, due to the additional attractive pull of standard gravity. As a result, the total CDM density perturbations given by
\begin{equation}
\delta _{\rm CDM} \equiv \frac{\Omega _{+}\delta _{+}}{\Omega _{\rm CDM}} + \frac{\Omega _{-}\delta _{-}}{\Omega _{\rm CDM}} \,
\label{delta_tot}
\end{equation}
will still grow in time with a rate that will significantly depend on the coupling strength and on the level of adiabaticity of the perturbations set.

\begin{figure*}
\includegraphics[scale=0.45]{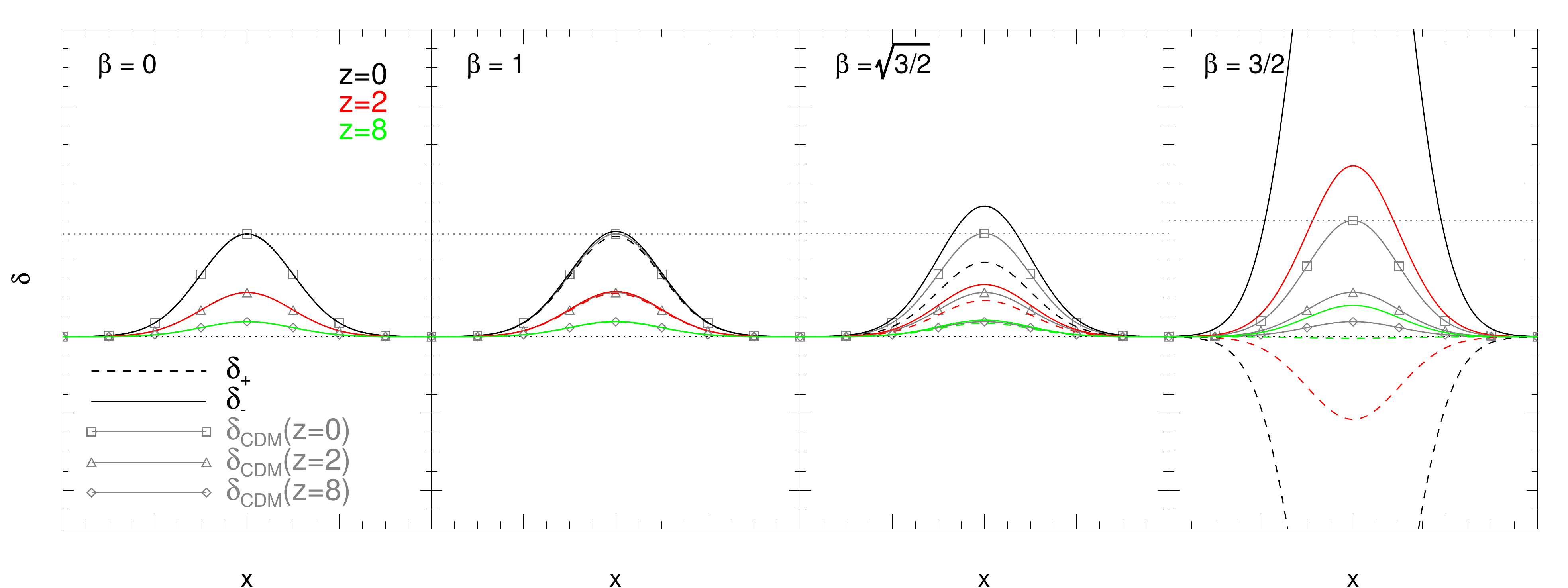}
\caption{The evolution of a gaussian density profile in real space for the positively- and negatively-coupled CDM species (dashed and solid lines, respectively) and for the total CDM density perturbation (grey solid lines) at redshifts $z=0\,,2\,,8$ (black, red, and green colors, respectively). The plot shows the overdensity profile $\delta (x)$ as a function of a radial comoving coordinate $x$ in arbitrary units. The grey squares, triangles, and diamonds are overplotted to highlight the shape of the total CDM density perturbation $\delta _{\rm CDM}$ at the three different redshifts considered in the plot. The horizontal dotted grey lines show the total amplitude of the CDM density perturbations at $z=0$. In the left panel, corresponding to an uncoupled case $\beta = 0$, no difference can be seen in the time evolution of the density profiles of the various components, while when progressively increasing the coupling value, moving to the right panels, one can start noticing a slight difference in the evolution of the positively- and negatively-coupled CDM species at low redshifts already for $\beta = 1$. The effect becomes more evident (and appears at earlier times) for larger values of the coupling, and for the most extreme case of $\beta = 3/2$ the two CDM species develop opposite density profiles by the present time, thereby showing the formation of isocurvature modes. Nevertheless, the total CDM perturbation amplitude is only mildly affected by this strikingly different evolution of the two individual CDM components.}
\label{fig:iso-adia}
\end{figure*}
\normalsize

Such peculiar behavior is well described by Fig.~\ref{fig:iso-adia}, where we show the evolution of a gaussian density perturbation
in real space in the context of four different McDE models with different coupling values. The plot displays the overdensity
profile $\delta (x)$ as a function of the radial comoving coordinate $x$ in arbitrary units. Different colors correspond to different redshifts (green for $z=8$, red for $z=2$, black for $z=0$) while dashed and solid lines correspond to the positively and negatively coupled species, respectively. The grey solid lines show the total CDM overdensity profile (\ref{delta_tot}) for the combined fluid, and the three redshifts are 
identified on these lines by different symbols (diamonds for $z=8$, triangles for $z=2$, and squares for $z=0$). For all the different cases we start
our integration at very high redshifts ($z\approx 10^{7}$) from an adiabatic configuration of the perturbations, such that $\Omega _{+}\delta _{+} = \Omega _{-}\delta _{-}$. As the plot shows, in the absence of coupling (left panel) the overdensity profiles in the two CDM species evolve in the same way, and the perturbations remain adiabatic until $z=0$. Correspondingly, the total CDM overdensity follows
the same evolution as the two individual species. On the contrary, in the presence of a non-vanishing coupling the evolutions of the two CDM species start to deviate from each other as time goes by. The negatively-coupled species starts to grow faster than its positively-coupled counterpart, due to the mutual repulsion given by the scalar fifth-force, thereby inducing the formation of isocurvature modes.
In particular, for the strongest coupling $\beta = 3/2$, we can clearly see in the rightmost panel of Fig.~\ref{fig:iso-adia} that already at high redshifts ($z=8$, green curves) the positively-coupled species shows no overdensity at all while the negatively-coupled one has a significantly more pronounced overdensity profile than in the uncoupled case. The situation evolves further at later times (red and black curves) with the positively-coupled species quickly developing an underdensity (i.e. a ``void") in correspondence to the large overdensity of the negatively-coupled one.

It is however very interesting to notice that, although the enhancement in the growth (or in the decay) rate of the individual CDM species dramatically increases with the coupling $\beta $, the evolution of the total CDM density perturbation results only mildly affected by the fifth-force, as one can see by comparing the amplitude of the overdensity profiles at $z=0$ in the different models, indicated by the horizontal dotted lines.
Unless for the most strongly coupled
case of $\beta = 3/2$, which exhibits a slight increase of the total amplitude at $z=0$, all the other models do not show any appreciable change in the evolution of the total CDM perturbation. This interesting result indicates that any probe of linear structure formation
based on the evolution of the total gravitational potential at linear scales will
not be able to distinguish a McDE scenario with a coupling as large as $\beta = \sqrt{3/2}$ (i.e. a model with a scalar fifth-force twice as strong as gravity) from a standard $\Lambda $CDM cosmology. In other words, despite the strikingly different evolution of the 
density fluctuations in the two CDM species and the appearance of isocurvature modes at late times, the superposition of the two
CDM fluids would effectively hide the DE-CDM interaction also at the level of linear perturbations, even for couplings of order unity and larger, thereby allowing for scalar forces of order gravity in the dark sector without any directly observational footprint.
As already discussed in \citet{Baldi_2012a}, this result shows that the claim by \citet{Brookfield_VanDeBruck_Hall_2008} that linear density perturbations allow to break the background degeneracy between McDE and $\Lambda $CDM models is actually valid only for relatively large coupling values.

It is therefore natural to try to extend the investigation of McDE models to the nonlinear regime
of structure formation, in order to study how their associated strong long-range scalar force
between CDM particle pairs, be it either attractive or repulsive, might affect the dynamics of collapsed objects at small scales,
and whether the nonlinear evolution of structures can provide some direct evidence of the presence of such dark interactions
that would remain otherwise completely hidden both at the background and linear perturbations levels.
To this end, we have run the first N-body simulations ever performed for this particular class of DE models, and we
will discuss their main features and results in the next Sections.

\section{The Simulations}
\label{sims}

In order to run our simulations, we have made use of the modified version by \citet{Baldi_etal_2010} of the widely used
parallel Tree-PM N-body code {\small GADGET} \citep[][]{gadget-2} that allows to self-consistently
simulate the evolution of structure formation processes in the context of generalized coupled DE cosmologies by
including all the relevant features of such models, from their specific background evolution to the mass variation of CDM
particles, to the effects of the fifth-force and of the extra friction acting on individual particles. Such modified code has been
widely used in the past to investigate coupled DE cosmologies with constant couplings \citep[][]{Baldi_etal_2010,Baldi_Viel_2010,Baldi_2011b} variable couplings \citep[][]{Baldi_2011a,Baldi_Pettorino_2011} and with bouncing
DE potentials \citep[][]{Baldi_2011c}, and has been recently employed for the development of the {\small CoDECS}\footnote{www.marcobaldi.it/research/CoDECS} Project \citep[][]{CoDECS} that provides publicly available numerical data from simulations of a wide range of coupled DE cosmologies. Here we apply for the first time the same code to the case of McDE, which does not require
any particular modification with respect to the original implementation since the code was developed in the first place to allow for 
multiple families of massive particles with individual couplings. Therefore, differently from previous applications where the two matter
families were identified with baryons and CDM (with the coupling of the former always set to zero) here we discard the baryons by 
switching off any hydrodynamical treatment of the corresponding particle type, and we identify the two matter families with the two different CDM species, simply setting the respective couplings to $\pm \beta$.

With this approach, we have run six cosmological N-body simulations of intermediate resolution for six different values of the coupling
($\beta = 0\,, 1/2\,, \sqrt{3}/2\,, 1\,, \sqrt{3/2}\,, 3/2$)
with the aim to highlight the main qualitative features of McDE models in the nonlinear regime. 
The simulations
follow the evolution of $2\times 256^{3}$ CDM particles in a periodic box of $100$ Mpc$/h$ aside, with a mass resolution at the starting
redshift $z_{i}=99$ of $m_{\pm } = 2.24\times 10^{9}$ M$_{\odot }/h$. The gravitational softening has been set to 
$\epsilon _{g} = 10$ kpc$/h$, corresponding to about $1/30$-th of the mean interparticle separation. 
Higher resolution simulations would have requested a considerably higher computational cost that is not justified by the needs
of the present preliminary investigation of McDE cosmologies, which is mainly aimed at sampling the parameter space of the modes
 and highlighting its most prominent features, and will be performed in future works.

Initial conditions have been
set by rescaling the amplitude of the linear power spectrum obtained with the Boltzmann code {\small CAMB} \citep[][]{camb} for a $\Lambda $CDM universe with WMAP7 cosmological parameters \citep[][]{wmap7} between last scattering ($z\approx 1100$) and the starting redshift of the simulations with the specific growth factor computed for each model by numerically integrating Eqs.~(\ref{gf_plus},\ref{gf_minus}). 
However, since the growth rate of McDE models is practically indistinguishable from that of $\Lambda $CDM at high redshifts, the initial conditions of all the six simulations are virtually identical. The random-phase realization of the linear
power spectrum in the initial conditions adopts the same random seed for all the models, such that the main shape
of the resulting large-scale structures will be identical in all the runs.

\section{Results}
\label{Results}

We now move to discuss the main outcomes of our numerical simulations, focusing on the
qualitative behavior of McDE cosmologies in the nonlinear regime of structure formation. 
As we will see in the remaining of the paper, our results show that only in this regime it is 
possible to identify some specific footprints that could allow to
detect and possibly constrain a long-range fifth-force of gravitational strength in the dark sector,
which would otherwise be completely hidden -- in these scenarios -- both at the level of the background and of the linear perturbations evolution.

\subsection{Large-scale structures in Multi-coupled Dark Energy cosmologies}

\begin{figure*}
\begin{minipage}{74mm}
\begin{center}
{\large $\beta = 0$}
\end{center}
\end{minipage}
\begin{minipage}{74mm}
\begin{center}
{\large $\beta = 1/2$}
\end{center}
\end{minipage}\\
\begin{minipage}{74mm}
\includegraphics[width=\linewidth]{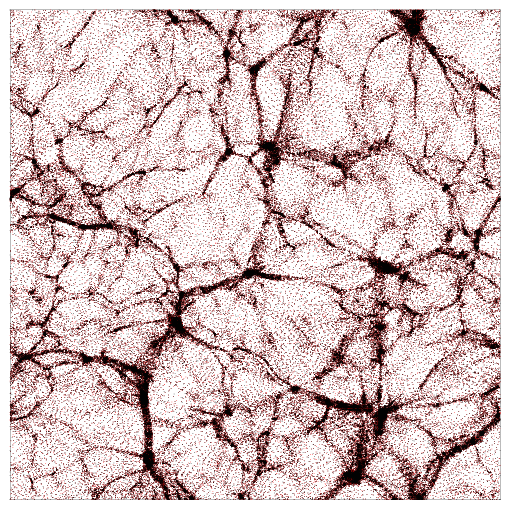}
\end{minipage}
\begin{minipage}{74mm}
\includegraphics[width=\linewidth]{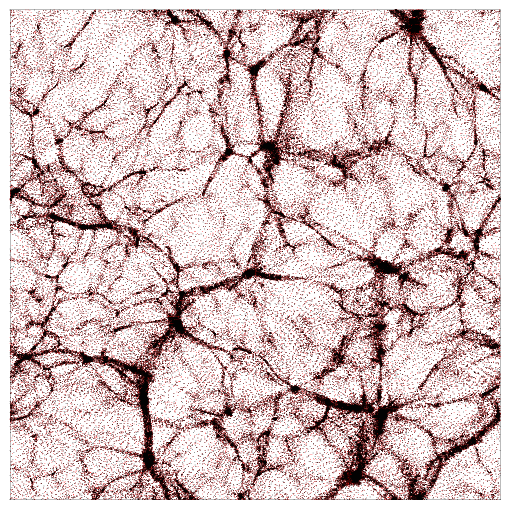}
\end{minipage}\\
\begin{minipage}{74mm}
\begin{center}
{\large $\beta = \sqrt{3}/2$}
\end{center}
\end{minipage}
\begin{minipage}{74mm}
\begin{center}
{\large $\beta = 1$}
\end{center}
\end{minipage}\\
\begin{minipage}{74mm}
\includegraphics[width=\linewidth]{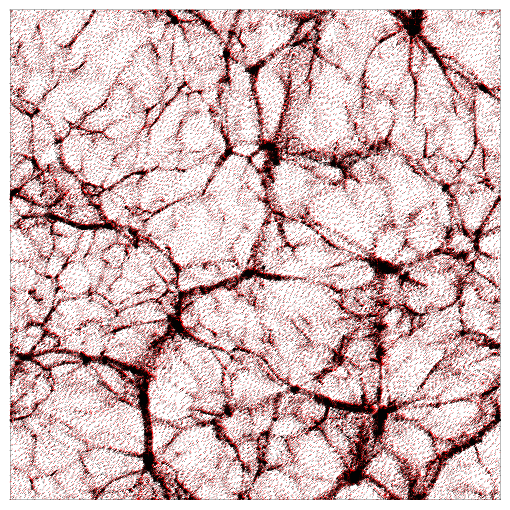}
\end{minipage}
\begin{minipage}{74mm}
\includegraphics[width=\linewidth]{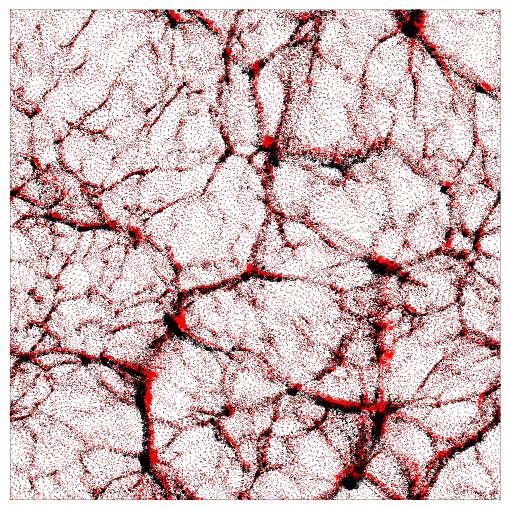}
\end{minipage}\\
\begin{minipage}{74mm}
\begin{center}
{\large $\beta = \sqrt{3/2}$}
\end{center}
\end{minipage}
\begin{minipage}{74mm}
\begin{center}
{\large $\beta = 3/2$}
\end{center}
\end{minipage}\\
\begin{minipage}{74mm}
\includegraphics[width=\linewidth]{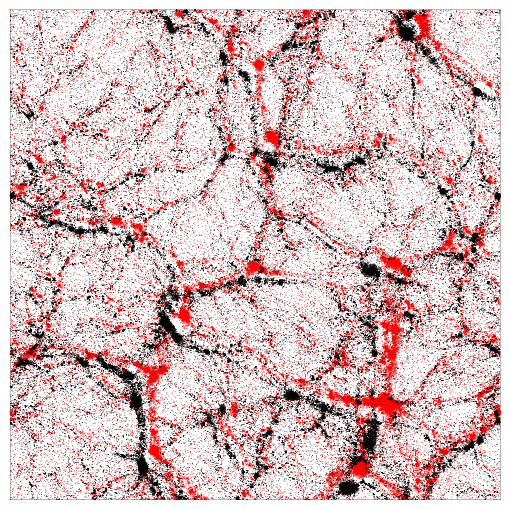}
\end{minipage}
\begin{minipage}{74mm}
\includegraphics[width=\linewidth]{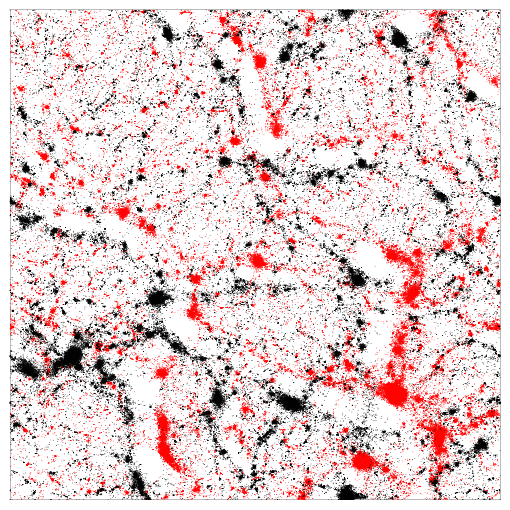}
\end{minipage}
\caption{The distribution of the positively- and negatively-coupled CDM particles (red and black points, respectively) in a slice of $100\times 100 \times 2$ Mpc$/h$ for different values of the coupling $\beta $. A higher-resolution color version of these plots is downloadable at: www.marcobaldi.it/data/MDM\_figures.tar}
\label{fig:slices}
\end{figure*}
\normalsize

We start the analysis of our simulations by investigating the qualitative shape and evolution of large-scale structures forming
from almost identical initial conditions
in the context of the different McDE models under investigation. In Figure~\ref{fig:slices} we plot the projected position
of particles of the positively- and negatively-coupled CDM species (as red and black points, respectively) in a slice of the cosmological box
of the different simulations at $z=0$. The thickness of the slice is $2$ Mpc$/h$ and the positively-coupled particles are plotted first such that
they might be covered by the subsequent plot of the negatively coupled particles. The different panels refer to different values of the coupling as explained in the
legend of the figure.

The left panel in the upper row displays the particle distribution in the absence of coupling ($\beta = 0$) and shows the familiar filamentary structure of
the cosmic web developing in any hierarchical structure formation scenario. In particular, the red points representing positively-coupled CDM particles are
barely visible (except in voids) as their spatial distribution closely follows that of the negatively-coupled particles represented by the black points: as expected, since both particle types
feel the same total force given only by the standard gravitational interaction, their distributions are indistinguishable from each other. Interestingly, the
situation does not seem to change appreciably also for a coupling value of $\beta = 1/2$ (right panel in the upper row), 
corresponding to a scalar fifth-force with a strength equal to one third of standard gravity, for which the qualitative shape of the large-scale structures appears indistinguishable
from the uncoupled case. This comparison already represents a very significant result in itself: a scalar fifth-force with one third the strength of standard gravity
does not seem to have any striking effect on the formation of large-scale cosmic structures even in the nonlinear regime probed with full N-body simulations. More detailed
investigations focused on the internal structure of collapsed objects might then be necessary to highlight possible observational footprints of scalar forces of this strength
in the context of McDE scenarios. We remind again here that a scalar fifth-force with strength $G/3$ between CDM particles is ruled out at more than $5 \sigma$ \citep[][]{Bean_etal_2008,Xia_2009,Baldi_Viel_2010} for standard coupled DE scenarios with only one CDM particle type.

If we now move to progressively larger couplings, however, some distinctive effects clearly start to appear, and the two CDM particle species do no longer look distributed
in space in the same way. In the middle row of Fig.~\ref{fig:slices}, we display the cases of $\beta = \sqrt{3}/2$ (left) and $\beta = 1$ (right). For the former model -- corresponding to  a fifth-force with the same strength as standard gravity -- the effect is still very mild and  the overall
shape of the cosmic web appears still identical to the uncoupled case, but a larger number of red points representing positively coupled particles starts to appear, showing
that the two CDM families are no longer identically distributed over the whole simulation box. The qualitative effect of the DE-CDM interaction becomes much more evident for a coupling of $\beta =1$, for which a clear shift between the two CDM species appears when looking at the edges of large and elongated structures (like e.g. cosmic filaments) 
showing an evident ongoing process of separation between the two CDM particle types.

Further increasing the coupling value, the bottom row of Fig.~\ref{fig:slices} shows in the left panel the particle distribution for $\beta = \sqrt{3/2}$,
where the development of two almost equally-shaped cosmic networks shifted from each other by a distance of a few Megaparsecs becomes apparent. This development
of ``mirror" structures in the cosmic web represents the first prominent feature that characterizes McDE models and that potentially allows to distinguish them
from a standard minimally coupled cosmology. Furthermore, it is possible to notice already by eye that the
distributions of the two individual CDM species also start showing some deviation from the overall shape of the CDM large-scale structure that was instead 
roughly preserved in the previous plots. In particular, for $\beta = \sqrt{3/2}$ -- corresponding to a fifth-force twice as strong as gravity -- one can clearly observe the fragmentation of large structures (like e.g. massive halos or filaments) into smaller and more peaked independent objects, and a significant depletion of cosmic voids. Finally, for a fifth-force
three times stronger than gravity, represented by the right plot of the bottom row of Fig.~\ref{fig:slices} ($\beta = 3/2$), the two CDM particle species are distributed in
a way that only vaguely retains the memory of the initial shape of the cosmic web, with isolated and very concentrated objects composed almost uniquely by one of the two different types of CDM particles. Very interestingly, the original shape of the large-scale structure can still be traced by following the extended voids that form in the regions  
between two ``mirror" structures made of different CDM species that have originated from the same ``seed" structure and that have subsequently departed from each other as a consequence of the strong repulsive force between the two particle families.

The visual inspection of the six panels of Fig.~\ref{fig:slices} then clearly shows how the nonlinear regime of structure formation can reveal distinctive features of 
McDE models that would be otherwise completely indistinguishable from $\Lambda $CDM in their background and linear perturbations evolution, and therefore
represents a unique handle to detect and possibly constrain the long-range scalar forces that arise in the context of these scenarios. In particular, this first qualitative analysis
shows that McDE models determine, for coupling values larger than unity, the formation and the progressive separation of ``mirror" cosmic structures mainly composed by particles of one single CDM species, and the corresponding onset of halo fragmentation processes at late times, when large ``seed" structures composed by a mixture of the
two CDM families split into smaller objects mainly composed by one single particle type. However, our results also show that for couplings smaller than unity, which include
also the case of a scalar force with the same strength of standard gravity, the impact of the fifth-force on the overall shape of large-scale structures is still very mild even at the nonlinear
level. 

\subsection{The nonlinear matter power spectrum}

We now move to a more quantitative analysis of the structures forming in the context of the different McDE models by
computing the nonlinear matter power spectrum of the different CDM species and of the total CDM density field directly from our simulations.
The power spectrum of these three different components is computed in Fourier space
by determining their individual density fields on a cartesian grid with the same size of the PM grid used for the long-range force computation
of the simulations (i.e. with $256$ grid nodes on each side) through a Cloud-in-Cell mass assignment algorithm. 
The nonlinear power spectra computed according to such procedure allow us to investigate the effects of the DE-CDM interaction
on the mass distribution up to the Nyquist frequency of the PM grid, corresponding to a wave number 
$k_{\rm Ny} \sim 8\, h/$Mpc, i.e. down to length scales relevant for present and upcoming weak lensing surveys.

\begin{figure*}
\includegraphics[scale=0.45]{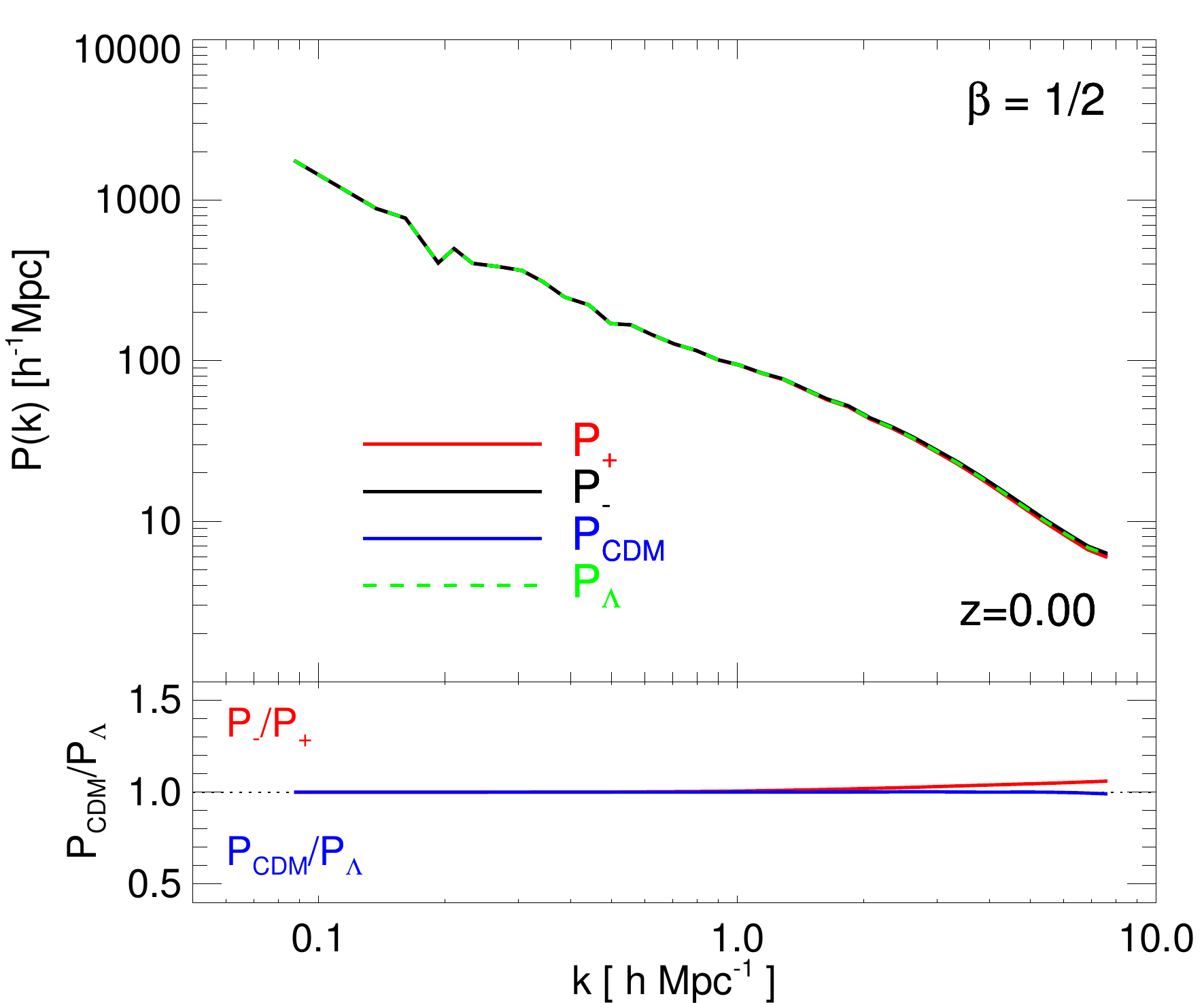}
\includegraphics[scale=0.45]{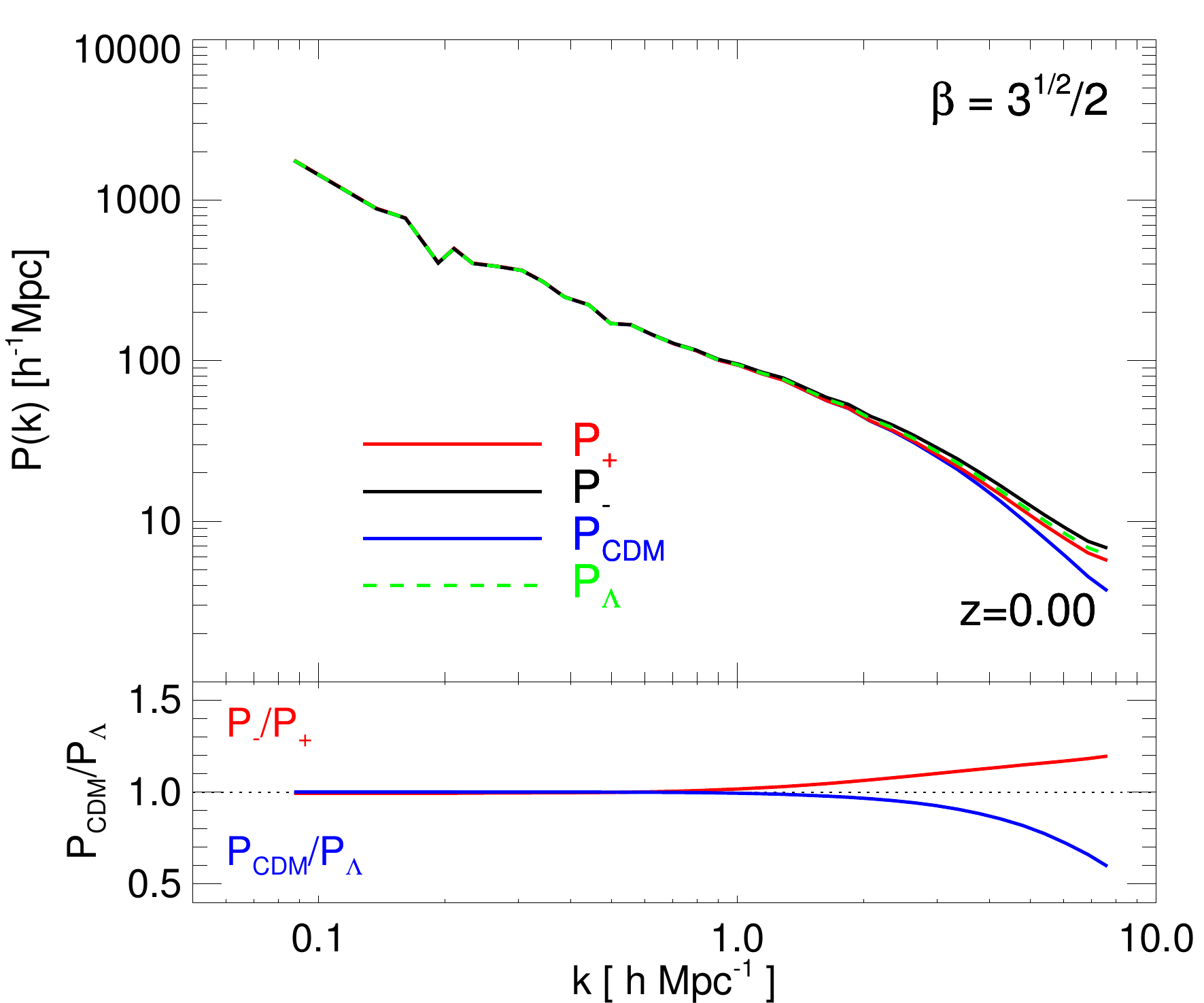}\\
\includegraphics[scale=0.45]{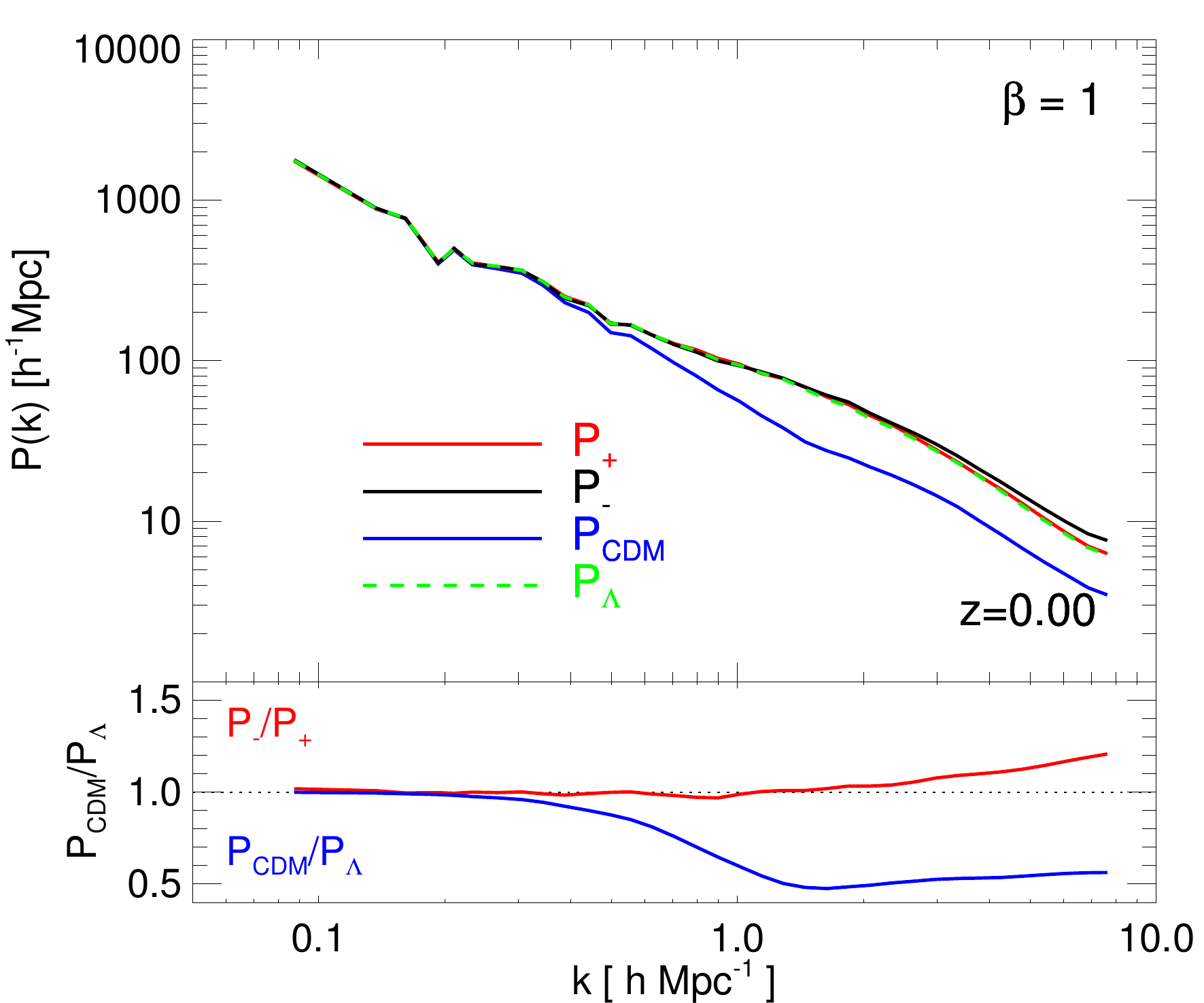}
\includegraphics[scale=0.45]{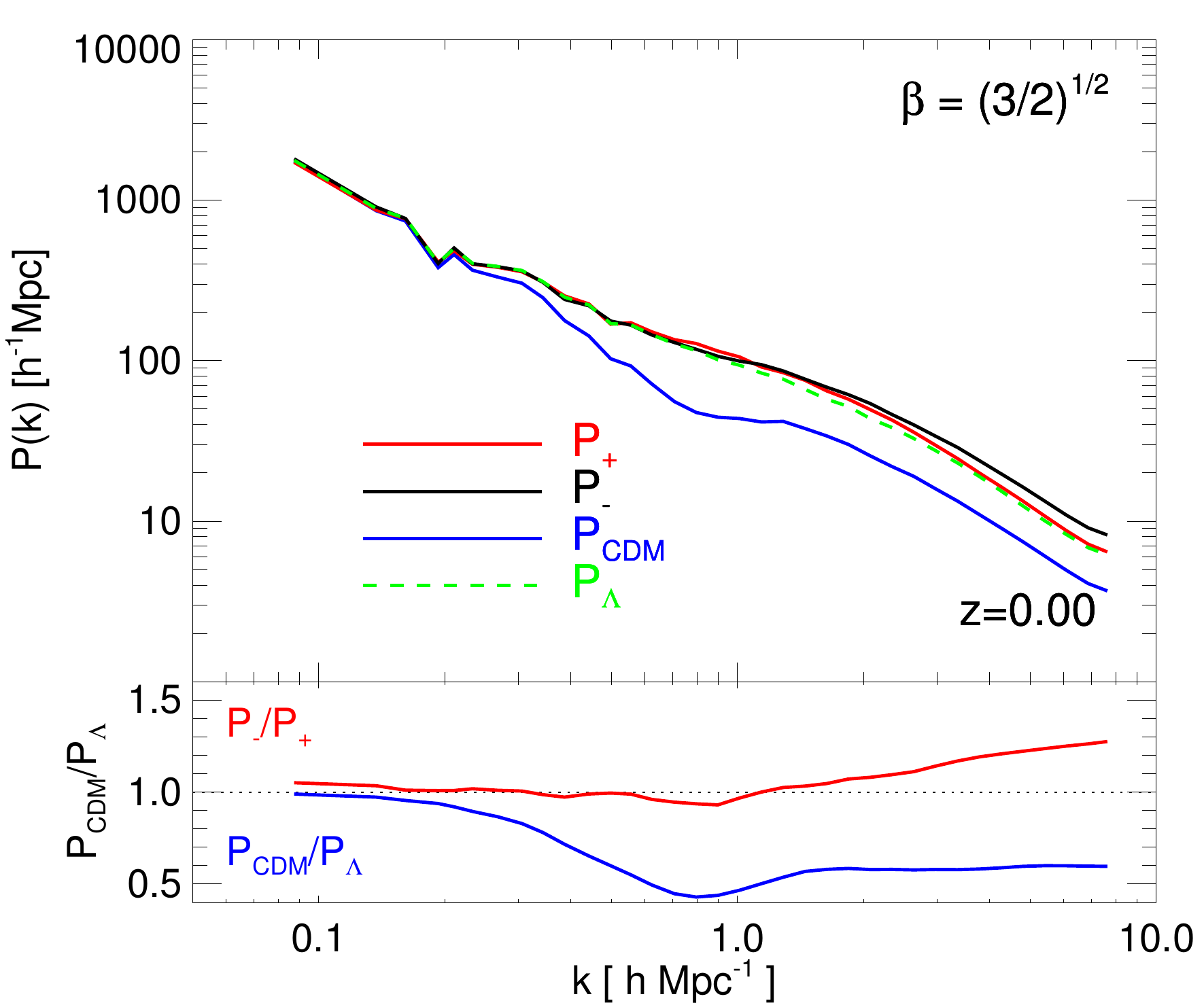}\\
\includegraphics[scale=0.45]{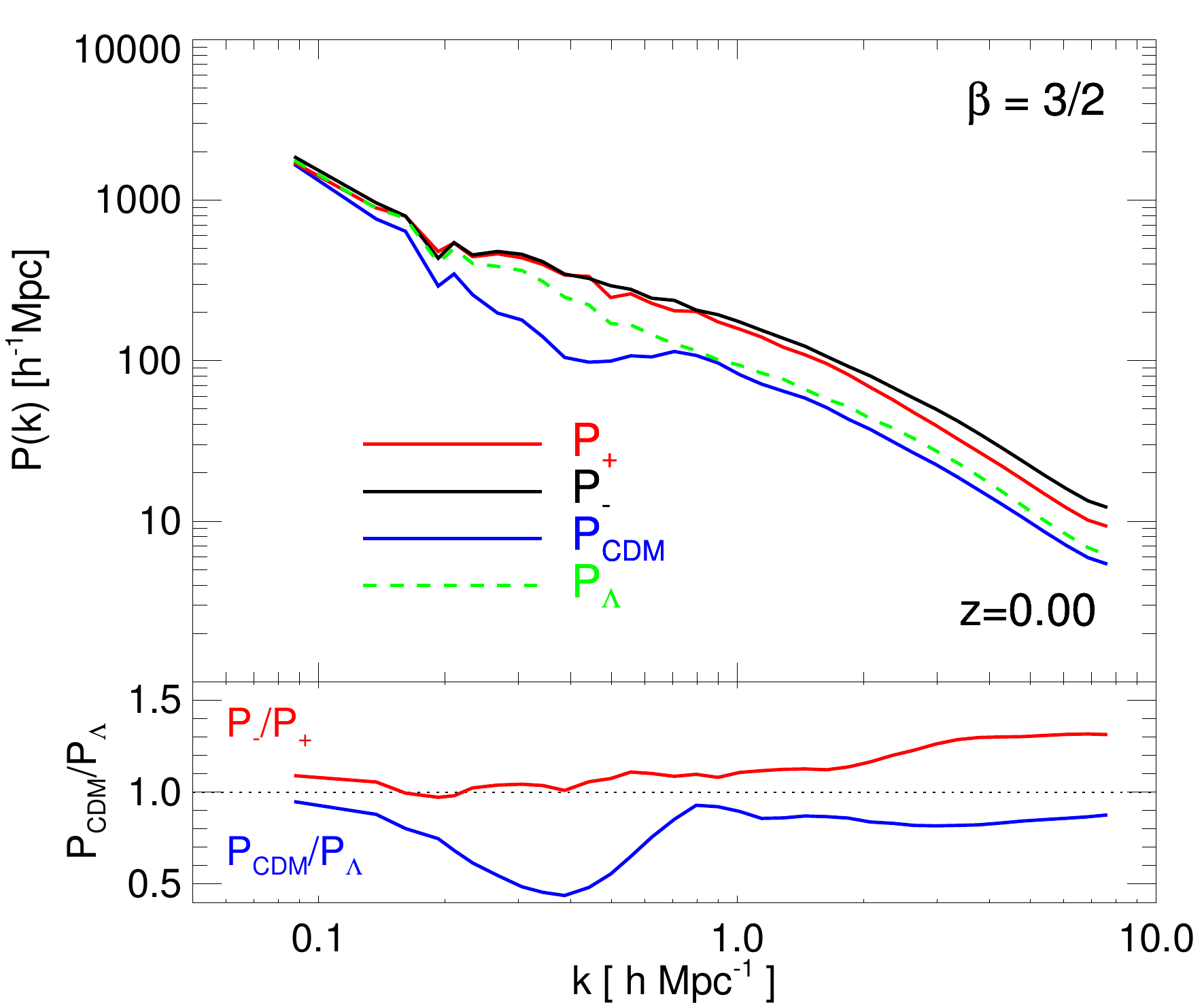}
\caption{The upper plot of each panel displays the matter power spectrum of the positively- and negatively-coupled CDM species $P_{+}(k)$ and $P_{-}(k)$ and of the total CDM component $P_{\rm CDM}(k)$, plotted as red, black, and blue solid lines, respectively, and the matter power spectrum of the standard $\Lambda $CDM cosmology overplotted for visual reference as a green dashed line, all at $z=0$. The bottom plots show in red the ratio of the power spectra of the two different CDM components $P_{-}(k)/P_{+}(k)$ and in blue the ratio of the total CDM power spectrum to the $\Lambda $CDM reference case, $P_{\rm CDM}/P_{\Lambda }$. As the different panels show, for a coupling of $\beta = 1/2$ no effect can be seen in the plots, while for larger coupling values a progressive suppression of power at small scales start to appear, which develops an almost-constant plateau at scales $k\gtrsim 1\, h/$Mpc for $\beta \geq 1$. Such plateau progressively moves back to the standard $\Lambda $CDM case for increasing couplings, and leaves a very characteristic maximum of the power suppression at intermediate scales between $k\sim 1\, h/$Mpc and $k\sim 4\, h/$Mpc.}
\label{fig:powerspectra}
\end{figure*}
\normalsize

In Figure~\ref{fig:powerspectra} we show for each model under investigation the nonlinear matter power spectra of the positively- and negatively-coupled
CDM species ($P_{+}(k)$ and $P_{-}(k)$, respectively plotted as red and black solid curves) as well as the power spectrum of the total CDM
density field ($P_{\rm CDM}(k)$, blue solid curve).
As a reference, we also overplot on each panel the nonlinear power spectrum $P_{\Lambda }(k)$ of the standard $\Lambda $CDM cosmology as a green dashed line.
The bottom plot of each panel shows in red the ratio between the power spectra of the two different CDM species, i.e. $P_{-}(k)/P_{+}(k)$, while in blue
it displays the ratio between the total CDM power spectrum of each specific model and the reference $\Lambda $CDM case, $P_{\rm CDM}(k)/P_{\Lambda }(k)$.

As one can see from the plots, for a coupling $\beta = 1/2$ the power spectra of the two different CDM species are practically indistinguishable from each other and from the total CDM one, and no difference can be seen also when comparing the latter to the standard $\Lambda $CDM power spectrum. This result reinforces the conclusion that, as suggested by the visual inspection of the large-scale structures discussed above, a scalar fifth-force with one third the strength of standard gravity does not affect in any appreciable way the density field in the context of McDE cosmologies down to the scales that we are able to resolve with present simulations. 

For larger coupling values, however, some effects start to clearly appear both in the relative evolution of the matter power spectra of the two different CDM species, and in the the total CDM power spectrum. In particular, for $\beta = \sqrt{3}/2$, corresponding to a scalar fifth-force with gravitational strength, 
a suppression of the total CDM power 
at scales smaller than $k\sim 1\, h/$Mpc  clearly appears in the plot, and progressively increases for smaller and smaller scales
up to a maximum suppression of $\sim 40\%$ at the Nyquist frequency, while for $k\gtrsim 1\, h/$Mpc the effect is basically absent and all the different power spectra are still identical. This is a particularly interesting result since it highlights a direct observational feature of a long-range scalar force with gravitational strength that appears only at scales that will be efficiently probed by the next generation of weak lensing surveys, while leaving larger scales as well as the background evolution of the universe completely unaffected. Another interesting feature of this plot is the different behavior at small scales of the two
CDM species, that confirms the evolution displayed in Fig.~\ref{fig:iso-adia} based on a purely linear treatment. More specifically, the plot shows how, starting from the smallest scales, perturbations that are initially adiabatic tend to develop some isocurvature component as a consequence of their different
gravitational interactions determined by the friction terms in Eqs.~(\ref{gf_plus},\ref{gf_minus}).

When moving to larger coupling values these effects become much more prominent, with a suppression of the total CDM power that can reach a level of about $40-50\%$ already at $k\sim 2\,h/$Mpc while still showing no suppression at all at the largest scales tested with our simulations.
It is nevertheless particularly interesting to notice that for couplings $\beta \geq 1$, the small-scale suppression of the matter power spectrum
features an almost constant plateau whose level tends to move back to the standard $\Lambda $CDM amplitude for increasing values of the coupling, 
leaving a clear maximum in the power spectrum suppression at intermediate scales. Such maximum (i.e. a minimum in the ratio $P_{\rm CDM}/P_{\Lambda }$ in Fig.~\ref{fig:powerspectra}) also moves to progressively larger scales
for increasing values of the coupling such that for the most extreme case addressed in this work, $\beta = 3/2$, the power spectrum is suppressed at the smallest scales 
by only about $10\%$ and  it still converges to the $\Lambda $CDM amplitude at the largest scales, while a maximum suppression of about $60\%$ appears at $k\sim 0.4\, h/$Mpc. This result is particularly interesting since it 
identifies another very distinctive observational footprint of McDE models, and 
shows an effect that could have only been tested through
full N-body simulations: the fragmentation of large collapsed objects made by a mixture of the two CDM particle types into smaller structures composed
primarily by a single particle type. 

This fragmentation is triggered by local deviations from perfect spherical symmetry of the density profiles of the two different CDM species within individual initial ``seed" overdensities. Such asymmetries arise as a consequence of tidal effects and merging processes, and then cannot be properly taken into account in the context of linear perturbation theory or even with a nonlinear spherical collapse treatment.
In fact, if the density profiles of the two CDM species within a common ``seed" overdensity are not both perfectly spherically symmetric,
the two species will start repelling each other away also along some radial direction, thereby triggering the fragmentation of the ``seed" structure into two smaller structures. 
Such fragmentation process then
determines an initial suppression of the nonlinear matter power, starting from the smallest scales which are the first ones to collapse, since the development of isocurvature modes and the separation of the sibling structures originated from an initial common ``seed"
reduce the total amplitude of density perturbations.
However, when two sibling halos which have originated from the same ``seed" structure are finally separated, they are no longer subject to the repulsive 
force given by the mixture of a comparable amount of particles of the two types, and only feel the strongly enhanced self-attraction given by
the sum of their standard self-gravity and their attractive fifth-force, thereby growing significantly faster than in the context of standard gravity and progressively becoming more overdense. This makes the small-scale power to grow faster than in $\Lambda $CDM 
in the ``post-fragmentation" phase, thereby progressively reducing the initial suppression. Such process of initial suppression and subsequent faster growth of the power spectrum amplitude is expected to occur at larger and larger scales
as time goes by, which also explains why the maximum suppression scale moves towards lower wavenumbers for increasing couplings. 

All these different peculiar features at scales potentially testable through galaxy clustering and weak lensing 
observations, and identified here for the first time, represent the only characteristic footprints of McDE cosmological models which might allow to directly constrain this type of scenarios and to possibly rule out long-range scalar forces in the dark sector, at least for sufficiently large interaction strengths.

\section{Conclusions}
\label{conclusions}

In this work we have presented the results of the first N-body simulations ever performed for Multi-coupled Dark Energy
cosmological models, characterized by the existence of two distinct CDM particle species interacting with opposite couplings with a classical
Dark Energy scalar field. Such models strongly alleviate the impact that standard coupled DE scenarios have on both the background expansion
history of the universe and the growth rate of linear density perturbations, and result completely indistinguishable from the standard $\Lambda $CDM
cosmology over a large range of coupling values. As a consequence, long-range scalar fifth-forces of gravitational strength that would be 
starkly incompatible with basic cosmological and astrophysical observations within standard coupled DE scenarios, are instead fully viable 
at the background and linear perturbations level in the context of Multi-coupled DE models.\\

We have therefore investigated for the first time the effects of these cosmologies in the nonlinear regime of structure formation, by means of a series
of intermediate-resolution N-body simulations that implement all the relevant effects characterizing Multi-coupled DE models, with the aim to qualitatively
highlight possible specific footprints of such scenarios and of their related long-range scalar forces over a wide range of parameter values. 
In particular, we have studied the qualitative evolution of the large-scale structure shape and of the relative spatial distribution of the two
different CDM families for different values of the DE-CDM coupling constant, and more quantitatively determined the impact of the different models
on the nonlinear matter power spectrum at scales that will be efficiently probed by present and future weak lensing surveys.\\

With the former analysis, we have shown how a scalar fifth-force with gravitational strength (or weaker) does not appreciably affect the overall shape 
of the large-scale structures forming in our simulated cosmological volumes, thereby making an extremely difficult task to distinguish Multi-coupled
DE models from minimally coupled scalar field cosmologies and even from the standard $\Lambda $CDM model. 
Higher-resolution simulations will be necessary in order to study the impact of such models on the inner regions of massive collapsed halos.
With larger values of the coupling,
instead, some peculiar features of the interaction can be identified in the nonlinear regime already at the present resolution level, for scenarios that would still appear completely indistinguishable 
from $\Lambda $CDM in the background and in the growth of linear perturbations. In particular, our simulations show that in Multi-coupeld DE models
the spatial distributions of the two different CDM species start to deviate from each other for sufficiently large coupling values, thereby leading to the
progressive separation of the two CDM families into two ``mirror" cosmic webs of collapsed halos and filaments that progressively move away from each other.
This peculiar process also leads to the fragmentation of large collapsed objects made of an almost equal mixture of the two CDM particle types
into smaller structures mainly composed by one particle species only. Such effects become dramatic for fifth-forces more than three times stronger
than standard gravity, and destroy the overall shape of the cosmic large-scale structures, thereby providing a direct way to constrain scenarios that would
be otherwise indistinguishable from $\Lambda $CDM.\\

For the latter and more quantitative analysis, we have computed the nonlinear matter power spectra of the two different CDM families and of the total
CDM fluid directly from our simulations, and shown that a potentially measurable suppression of power at scales $k\gtrsim 1\, h/$Mpc starts to appear
already for scalar fifth-forces with the same strength as standard gravity, and progressively becomes more significant for larger values of the DE-CDM
coupling. Finally, although such small-scale suppression tends to decrease again for even larger coupling values, 
as a consequence of the enhanced growth of the separate ``mirror" halos once they have departed from each other, a characteristic 
maximum suppression at intermediate scales clearly remains imprinted in the matter power spectrum and represents a very characteristic footprint
that could allow to rule out such extremely large coupling values using future weak lensing surveys.\\

We can therefore conclude that McDE models -- which are completely indistinguishable from the standard $\Lambda $CDM cosmology both
at the background and at the linear perturbations level for a wide range of coupling values -- can have interesting and peculiar footprints in the nonlinear regime of structure formation. However, such effects become particularly prominent, and might possibly rule out or constrain the model, only for relatively large values of the dimensionless coupling,
while for scalar interactions up to gravitational strength they remain sufficiently subtle to evade all the possible diagnostics that are allowed by the resolution of present simulations. A more detailed investigation of these models is therefore required to explore their possible effects on the strongly nonlinear systems represented by the internal 
regions of massive collapsed halos, and to provide a more quantitative determination of their observational signatures.

\section*{Acknowledgments}

I am deeply thankful to Luca Amendola for useful discussions on the models.
This work has been supported by 
the DFG Cluster of Excellence ``Origin and Structure of the Universe''and by
the TRR33 Transregio Collaborative Research Network on the ``Dark%
Universe''.
%*****************************************************************************
\bibliographystyle{mnras}
\bibliography{baldi_bibliography}

\begin{thebibliography}{65}
\expandafter\ifx\csname natexlab\endcsname\relax\def\natexlab#1{#1}\fi

\bibitem[Aarssen et~al.(2012)Aarssen, Bringmann \&
  Pfrommer]{Aarssen_Bringmann_Pfrommer_2012}
Aarssen L. G.~d., Bringmann T., Pfrommer C., 2012, arXiv:1205.5809

\bibitem[Amendola(2000)]{Amendola_2000}
Amendola L., 2000, Phys. Rev., D62, 043511

\bibitem[Amendola(2004)]{Amendola_2004}
Amendola L., 2004, Phys. Rev., D69, 103524

\bibitem[{Amendola} et~al.(2012){Amendola}, {Appleby}, {Bacon}
  et~al.]{Euclid_TWG}
{Amendola} L., {Appleby} S., {Bacon} D., et~al., 2012, eprint arXiv:1206.1225

\bibitem[Amendola et~al.(2011)Amendola, Pettorino, Quercellini \&
  Vollmer]{Amendola_Pettorino_Quercellini_2012}
Amendola L., Pettorino V., Quercellini C., Vollmer A., 2011, arXiv:1111.1404

\bibitem[Amendola \& Quercellini(2003)]{Amendola_Quercellini_2003}
Amendola L., Quercellini C., 2003, Phys. Rev., D68, 023514

\bibitem[Appleby \& Weller(2010)]{Appleby_Weller_2010}
Appleby S.~A., Weller J., 2010, JCAP, 1012, 006

\bibitem[Armendariz-Picon et~al.(2001)Armendariz-Picon, Mukhanov \&
  Steinhardt]{kessence}
Armendariz-Picon C., Mukhanov V.~F., Steinhardt P.~J., 2001, Phys. Rev., D63,
  103510

\bibitem[Astier et~al.(2006)]{SNLS}
Astier P., et~al., 2006, Astron. Astrophys., 447, 31

\bibitem[Astier \& Pain(2012)]{Astier_Pain_2012}
Astier P., Pain R., 2012, arXiv:1204.5493

\bibitem[Baccigalupi et~al.(2000)Baccigalupi, Matarrese \&
  Perrotta]{Baccigalupi_Matarrese_2000}
Baccigalupi C., Matarrese S., Perrotta F., 2000, Phys. Rev., D62, 123510

\bibitem[{Baldi}(2011{\natexlab{a}})]{Baldi_2011b}
{Baldi} M., 2011{\natexlab{a}}, \mnras, 414, 116

\bibitem[{Baldi}(2011{\natexlab{b}})]{Baldi_2011a}
{Baldi} M., 2011{\natexlab{b}}, \mnras, 411, 1077

\bibitem[{Baldi}(2012{\natexlab{a}})]{Baldi_2011c}
{Baldi} M., 2012{\natexlab{a}}, \mnras, 420, 430

\bibitem[{Baldi}(2012{\natexlab{b}})]{Baldi_2012a}
{Baldi} M., 2012{\natexlab{b}}, arXiv:1204.0514

\bibitem[{Baldi}(2012{\natexlab{c}})]{CoDECS}
{Baldi} M., 2012{\natexlab{c}}, \mnras, 422, 1028

\bibitem[{Baldi} \& {Pettorino}(2011)]{Baldi_Pettorino_2011}
{Baldi} M., {Pettorino} V., 2011, Mon. Not. Roy. Astron. Soc., 412, L1

\bibitem[{Baldi} et~al.(2010){Baldi}, {Pettorino}, {Robbers} \&
  {Springel}]{Baldi_etal_2010}
{Baldi} M., {Pettorino} V., {Robbers} G., {Springel} V., 2010, \mnras, 403,
  1684

\bibitem[{Baldi} \& Viel(2010)]{Baldi_Viel_2010}
{Baldi} M., Viel M., 2010, Mon. Not. Roy. Astron. Soc., 409, 89

\bibitem[Bartelmann(2010)]{Bartelmann_2010}
Bartelmann M., 2010, Rev.Mod.Phys., 82, 331

\bibitem[Bean et~al.(2008)Bean, Flanagan, Laszlo \& Trodden]{Bean_etal_2008}
Bean R., Flanagan E.~E., Laszlo I., Trodden M., 2008, Phys. Rev., D78, 123514

\bibitem[Bergstrom(2012)]{Bergstrom_2012}
Bergstrom L., 2012, arXiv:1205.4882

\bibitem[Bertone et~al.(2005)Bertone, Hooper \& Silk]{Bertone_Hooper_Silk_2005}
Bertone G., Hooper D., Silk J., 2005, Phys. Rept., 405, 279

\bibitem[Bertotti et~al.(2003)Bertotti, Iess \&
  Tortora]{Bertotti_Iess_Tortora_2003}
Bertotti B., Iess L., Tortora P., 2003, Nature, 425, 374

\bibitem[Brans \& Dicke(1961)]{Brans_Dicke_1961}
Brans C., Dicke R., 1961, Phys.Rev., 124, 925

\bibitem[Brookfield et~al.(2008)Brookfield, van~de Bruck \&
  Hall]{Brookfield_VanDeBruck_Hall_2008}
Brookfield A.~W., van~de Bruck C., Hall L. M.~H., 2008, Phys. Rev., D77, 043006

\bibitem[Caldera-Cabral et~al.(2009)Caldera-Cabral, Maartens \&
  Urena-Lopez]{CalderaCabral_2009}
Caldera-Cabral G., Maartens R., Urena-Lopez L.~A., 2009, Phys. Rev., D79,
  063518

\bibitem[Clemson et~al.(2011)Clemson, Koyama, Zhao, Maartens \&
  Valiviita]{Clemson_etal_2011}
Clemson T., Koyama K., Zhao G.-B., Maartens R., Valiviita J., 2011,
  arXiv:1109.6234, * Temporary entry *

\bibitem[Damour et~al.(1990)Damour, Gibbons \&
  Gundlach]{Damour_Gibbons_Gundlach_1990}
Damour T., Gibbons G.~W., Gundlach C., 1990, Phys. Rev. Lett., 64, 123

\bibitem[Deffayet et~al.(2002)Deffayet, Dvali, Gabadadze \&
  Vainshtein]{Deffayet_etal_2002}
Deffayet C., Dvali G., Gabadadze G., Vainshtein A.~I., 2002, Phys.Rev., D65,
  044026

\bibitem[{Farrar} \& {Peebles}(2004)]{Farrar2004}
{Farrar} G.~R., {Peebles} P.~J.~E., 2004, \apj, 604, 1

\bibitem[Fu et~al.(2008)]{Fu_etal_2008}
Fu L., et~al., 2008, Astron. Astrophys., 479, 9

\bibitem[Giannantonio et~al.(2008)]{Giannantonio_etal_2008}
Giannantonio T., et~al., 2008, Phys. Rev., D77, 123520

\bibitem[Guzzo et~al.(2008)Guzzo, Pierleoni, Meneux, Branchini, Fevre
  et~al.]{Guzzo_etal_2008}
Guzzo L., Pierleoni M., Meneux B., Branchini E., Fevre O.~L., et~al., 2008,
  Nature, 451, 541

\bibitem[Hellwing et~al.(2010)Hellwing, Knollmann \& Knebe]{Hellwing_etal_2010}
Hellwing W.~A., Knollmann S.~R., Knebe A., 2010, arXiv:1004.2929

\bibitem[Hinterbichler \& Khoury(2010)]{Hinterbichler_Khoury_2010}
Hinterbichler K., Khoury J., 2010, Phys.Rev.Lett., 104, 231301

\bibitem[Keselman et~al.(2010)Keselman, Nusser \&
  Peebles]{Keselman_Nusser_Peebles_2010}
Keselman J.~A., Nusser A., Peebles P., 2010, Phys.Rev., D81, 063521

\bibitem[Khlopov(1995)]{Khlopov_1995}
Khlopov M.~Y., 1995,  133--138

\bibitem[Khoury \& Weltman(2004)]{Khoury_Weltman_2004}
Khoury J., Weltman A., 2004, Phys.Rev., D69, 044026

\bibitem[Komatsu et~al.(2011)]{wmap7}
Komatsu E., et~al., 2011, Astrophys. J. Suppl., 192, 18

\bibitem[Koyama et~al.(2009)Koyama, Maartens \& Song]{Koyama_etal_2009}
Koyama K., Maartens R., Song Y.-S., 2009, JCAP, 0910, 017

\bibitem[La~Vacca et~al.(2009)La~Vacca, Kristiansen, Colombo, Mainini \&
  Bonometto]{LaVacca_etal_2009}
La~Vacca G., Kristiansen J.~R., Colombo L. P.~L., Mainini R., Bonometto S.~A.,
  2009, JCAP, 0904, 007

\bibitem[Lewis et~al.(2000)Lewis, Challinor \& Lasenby]{camb}
Lewis A., Challinor A., Lasenby A., 2000, Astrophys. J., 538, 473

\bibitem[Li \& Barrow(2011)]{Li_Barrow_2011}
Li B., Barrow J.~D., 2011, Phys. Rev., D83, 024007

\bibitem[Loeb \& Weiner(2011)]{Loeb_Weiner_2011}
Loeb A., Weiner N., 2011, Phys. Rev. Lett., 106, 171302

\bibitem[Macci\`{o} et~al.(2004)Macci\`{o}, Quercellini, Mainini, Amendola \&
  Bonometto]{Maccio_etal_2004}
Macci\`{o} A.~V., Quercellini C., Mainini R., Amendola L., Bonometto S.~A.,
  2004, Phys. Rev., D69, 123516

\bibitem[{Maccio'} et~al.(2012){Maccio'}, {Ruchayskiy}, {Boyarsky} \&
  {Munoz-Cuartas}]{Maccio_etal_2012}
{Maccio'} A.~V., {Ruchayskiy} O., {Boyarsky} A., {Munoz-Cuartas} J.~C., 2012,
  arXiv:1202.2858

\bibitem[Mantz et~al.(2010)Mantz, Allen, Rapetti \& Ebeling]{Mantz_etal_2010}
Mantz A., Allen S.~W., Rapetti D., Ebeling H., 2010, Mon. Not. Roy. Astron.
  Soc., 406, 1759

\bibitem[Percival et~al.(2001)]{Percival_etal_2001}
Percival W.~J., et~al., 2001, Mon. Not. Roy. Astron. Soc., 327, 1297

\bibitem[Percival et~al.(2010)]{Percival_etal_2010}
Percival W.~J., et~al., 2010, Mon. Not. Roy. Astron. Soc., 401, 2148

\bibitem[Perlmutter et~al.(1999)]{Perlmutter_etal_1999}
Perlmutter S., et~al., 1999, Astrophys. J., 517, 565

\bibitem[Perrotta et~al.(2000)Perrotta, Baccigalupi \&
  Matarrese]{Perrotta_etal_2000}
Perrotta F., Baccigalupi C., Matarrese S., 2000, Phys. Rev., D61, 023507

\bibitem[Pettorino \& Baccigalupi(2008)]{Pettorino_Baccigalupi_2008}
Pettorino V., Baccigalupi C., 2008, Phys. Rev., D77, 103003

\bibitem[Pettorino et~al.(2005)Pettorino, Baccigalupi \&
  Mangano]{Pettorino_etal_2005}
Pettorino V., Baccigalupi C., Mangano G., 2005, JCAP, 0501, 014

\bibitem[Ratra \& Peebles(1988)]{Ratra_Peebles_1988}
Ratra B., Peebles P. J.~E., 1988, Phys. Rev., D37, 3406

\bibitem[Riess et~al.(1998)]{Riess_etal_1998}
Riess A.~G., et~al., 1998, Astron. J., 116, 1009

\bibitem[Schmidt et~al.(1998)]{Schmidt_etal_1998}
Schmidt B.~P., et~al., 1998, Astrophys.J., 507, 46

\bibitem[Schrabback et~al.(2010)Schrabback, Hartlap, Joachimi, Kilbinger, Simon
  et~al.]{Schrabback_etal_2010}
Schrabback T., Hartlap J., Joachimi B., Kilbinger M., Simon P., et~al., 2010,
  Astron.Astrophys., 516, A63

\bibitem[Sherwin et~al.(2011)Sherwin, Dunkley, Das, Appel, Bond
  et~al.]{Sherwin_etal_2011}
Sherwin B.~D., Dunkley J., Das S., Appel J.~W., Bond J.~R., et~al., 2011,
  Phys.Rev.Lett., 107, 021302

\bibitem[Springel(2005)]{gadget-2}
Springel V., 2005, Mon. Not. Roy. Astron. Soc., 364, 1105

\bibitem[Vainshtein(1972)]{Vainshtein_1972}
Vainshtein A., 1972, Phys.Lett., B39, 393

\bibitem[Wetterich(1988)]{Wetterich_1988}
Wetterich C., 1988, Nucl. Phys., B302, 668

\bibitem[Wetterich(1995)]{Wetterich_1995}
Wetterich C., 1995, Astron. Astrophys., 301, 321

\bibitem[Will(2005)]{Will_2005}
Will C.~M., 2005, Living Rev.Rel., 9, 3

\bibitem[Xia(2009)]{Xia_2009}
Xia J.-Q., 2009, Phys. Rev., D80, 103514

\end{thebibliography}

\label{lastpage}

\end{document}